\documentclass[aps,preprint,showpacs,preprintnumbers,amsmath,amssymb,prl]{revtex4}
\usepackage{epsf,amsmath,amssymb,verbatim,color,multirow,pifont}
\usepackage{graphicx}
\begin{document}

\newcommand{\EQ}{Eq.~}
\newcommand{\EQS}{Eqs.~}
\newcommand{\FIG}{Fig.~}
\newcommand{\FIGS}{Figs.~}
\newcommand{\SEC}{Sec.~}
\newcommand{\SECS}{Secs.~}

\title{Diversity and critical behavior in prisoner's dilemma game}
\author{C.-K. Yun$^1$, N.~Masuda$^{2,3}$, and B.~Kahng$^{1,4}$}
\address{{$^1$Department of Physics and Astronomy, Seoul National
University, Seoul 151-747, Korea}\\
{$^2$ Department of Mathematical Informatics, The
University of Tokyo, Tokyo 113-8656, Japan}\\
{$^3$ PRESTO, Japan Science and Technology Agency, 4-1-8 Honcho, Kawaguchi, Saitama 332-0012, Japan}\\
{$^4$ School of Physics, Korea Institute for Advanced Study, Seoul
130-722, Korea}}
\date{\today}

\begin{abstract}
The prisoner's dilemma (PD) game is a simple model for understanding cooperative
patterns in complex systems consisting of selfish individuals.
Here, we study a PD game problem in scale-free networks containing
hierarchically organized modules and controllable shortcuts connecting separated hubs.
We find that cooperator clusters exhibit a percolation transition in
the parameter space $(p,b)$, where $p$ is the occupation probability of shortcuts
and $b$ is the temptation payoff in the PD game. The cluster size distribution follows a power
law at the transition point. Such a critical behavior, resulting from the
combined effect of stochastic processes in the PD game and the heterogeneous
structure of complex networks, illustrates the diversity of social relationships
and the self-organization of cooperator communities in real-world systems.
\end{abstract}

\pacs{89.75.Hc, 89.75.Da, 89.75.Fb}
\maketitle

Social interactions between individuals are often cooperative or competitive,
and certain emerging patterns in these interactions can be recognized.
In particular, an understanding of social dilemmas where
the optimal strategy for the whole society is not the same as that for
each individual has been an attractive topic of interdisciplinary research.
The prisoner's dilemma (PD) game has been used as a basic model in understanding
the patterns that emerge in the case of social dilemmas \cite{Axelrod84book,Nowak06book}.
In the PD game, which was originally designed as a two-player game,
a player earns a larger payoff by unilateral defection than by mutual
cooperation. Therefore, even if both players attain nothing when
they defect, the optimal choice would be defection. On the other hand,
in real systems, people are cooperative and altruistic. Thus, it will be
interesting to resolve such a paradox by using a simple model.

A social system is more accurately described by a
network of players \cite{net reviews}. If players in a network
change their strategies according to evolutionary dynamics
(i.e., by imitating successful neighbors), we often observe mutual
cooperation, and the system exhibits diverse patterns
\cite{Nowak06book,SzaboFath07}. These diverse patterns result
from the collective dynamics of the players' interactions with each
other~\cite{diversity}.
While such diverse patterns are present and probably
functional in social relationships, their diversity has not
been clearly understood yet. In this Letter, we investigate the
origin of the formation of such diverse patterns by using the PD game.

The enhanced cooperation and emerging patterns in the PD game were first clarified
in the Euclidean space \cite{Axelrod84book,spatial reciprocity}.
Recently, the PD game was extended to complex heterogeneous
networks such as scale-free (SF) networks, in which
the number of the nearest neighbors of a node (which is known as the degree
of the node) follows a power law. In random SF networks such as the
Barab\'asi and Albert model~\cite{Barabasi99}, the mean distance between two nodes
scales logarithmically with the system size. In such small-world SF networks,
the density of cooperators is significantly higher \cite{Santos-Duran}
than that in the case of other networks, including those in the Euclidean space.
Once cooperation is stabilized at the hubs at an early stage of dynamics,
cooperation spreads to nodes with smaller degrees~\cite{Santos-Duran,Gomez-Poncela}.
Thus, the hub plays a crucial role in spreading cooperation.

Many SF networks in the real world are not as random as that in
the Barab\'asi and Albert model; they contain modular structure.
Moreover, the modular structure is hierarchically organized~\cite{hierarchy}.
In such modular SF networks, hubs are separated from each other,
and the mean distance between two nodes often scales in a power-law
manner with the system size \cite{fractal nets}.
Such networks are called large-world or fractal networks. Cooperation
would in such networks is less than that in random SF networks,
because the large distance between hubs generally reduces cooperation
~\cite{Santos-Duran,Gomez-Poncela}.

\begin{figure}
\begin{center}
\includegraphics[width=9cm]{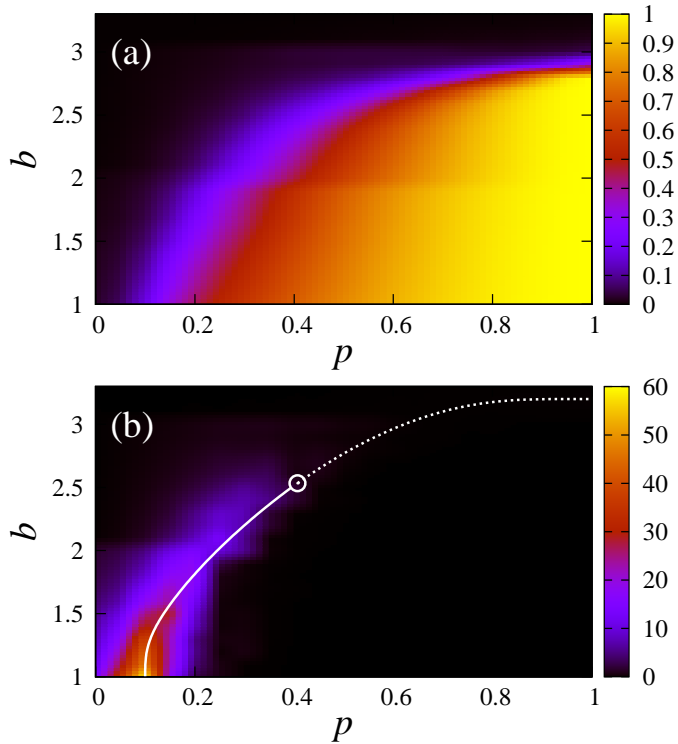}
\caption{(Color online) (a) Giant cluster size in the parameter
space of $(p,b)$ for the hierarchical network with system size $N=10,924$.
Data are averaged over 100 network configurations.
(b) Susceptibility under the same condition as that in the case of (a).
The solid and dotted curves are the loci along the peaks of susceptibility:
the peaks represent continuous and discontinuous percolation transitions, respectively.}~\label{fig:op}
\end{center}
\end{figure}

Social networks in the real world are often at the boundary between small-world
and large-world networks~\cite{Deokjae}. Thus, in this Letter, we consider the
PD game in artificial networks in which the number of edges between
the separated hubs is controlled by the occupation probability $p$ and examine
the patterns in the interactions between the cooperators as the network
transforms from a large-world to a small-world networks as $p$ increases.
We find that clusters composed of cooperators undergo a percolation transition
in the parameter space of $(p,b)$, where $b$ is the temptation payoff in the PD game.
Interestingly, the percolation transition occurs either
continuously or discontinuously as $p$ is increased for a fixed $b$
or $b$ is decreased for a fixed $p$. Therefore, there exists a
tricritical-like point $(p_t,b_t)$ such that for a fixed
$b < b_t$ ($b > b_t$) or $p< p_t$ ($p > p_t$),
the giant cluster of cooperators grows continuously (discontinuously) (Fig.~\ref{fig:op}(a)).
The phase diagram is shown in Fig.~\ref{fig:op}(b).
Furthermore, the size distribution of the cooperator clusters exhibits
a power-law behavior near the percolation threshold as long as $p < p_t$.
This emerging pattern suggests that the cooperators self-organize their
clusters to attain a critical state, which enhances the diversity within the
system.

On the basis of previous papers \cite{SzaboFath07,Santos-Duran,Gomez-Poncela},
we define the payoff matrix by
\begin{equation}
\bordermatrix{
 & C & D \cr
C & 1 & 0 \cr
D & b & 0 \cr}. \;
\label{eq:payoff matrix}
\end{equation}
A row player selects one of two states, either cooperation (C)
or defection (D), so does a column player. Each entry in
\EQ\eqref{eq:payoff matrix} represents the payoff that the row
player obtains. If the two players start with mutual cooperation, both players are tempted to defect in order to obtain a larger payoff (temptation) i.e., $b > 1$. If both players act selfishly and choose to defect, both obtain no payoff; this is the unique Nash equilibrium
of the game. However, the individual rewards for both the players could be higher if they
mutually cooperate.

We examine the evolutionary PD game for several types of complex networks with
equal initial densities of C and D. At each time step (or
round), each player $i$ interacts with all of its $k_i$ neighbors; here,
$k_i$ is the degree of the node $i$. Player $i$'s payoff in one round $P_i$
is the sum of all the payoffs earned by playing against the $k_i$ neighbors.
Player $i$ updates its strategy according to the
following rule~\cite{Santos-Duran}: A neighbor $j$ of player $i$ is chosen
with equal probability $1/k_i$. Then, if $P_j > P_i$, $i$ copies $j$'s strategy
with probability $\left(P_j -
P_i\right)\big/\left[b \max\left(k_i,k_j\right)\right]$. The
denominator normalizes the probability such that the probability is
between 0 and 1. On the other hand, if $P_j < P_i$,  $i$ does not change
its strategy. The rule for updating strategies is synchronously applied
by all the players. This procedure is repeated in subsequent rounds.

In each round, a player chooses either C or D. However,
in long-term dynamics, players are categorized as~\cite{Gomez-Poncela}:
a permanent cooperator, a permanent defector, or an unstable player that
continuously changes its state between C and D. We numerically simulate the PD game
until the density of the cooperators does not change with time.
We simulate the PD game for up to $2 \times 10^4$ rounds after a steady state is reached
for each players' configuration at initial. A permanent cooperator (defector) is defined as
the node that chooses only C (D) for the last $10^4$ rounds. The rest of the nodes
that do not satisfy this condition are regarded as unstable players. The density
of the permanent cooperators is denoted by $\rho_c$.

\begin{figure}
\begin{center}
\includegraphics[width=9cm]{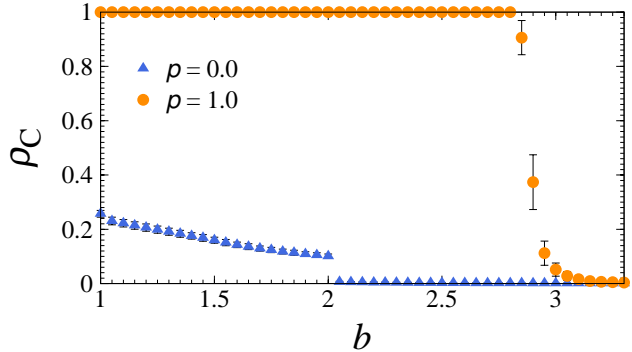}
\caption{(Color online) Density of cooperators as a function of temptation
payoff $b$ for the two extreme cases, $p=0$ and $p=1$ in the
hierarchical model.}
\label{fig:coop}
\end{center}
\end{figure}

The PD game is played on the so-called hierarchical network,
introduced in~\cite{Hinczewski06}. This network is constructed by
repeating a simple structural mapping as follows: We begin with an
edge, which is replaced with a diamond in the next iteration step.
A horizontal edge between two nodes at the diagonal positions is
present with probability $p$. This process is repeated for each
edge in the diamond (except the horizontal bond) in subsequent
iterations until the obtained network attains the desired system
size. This network is scale-free and has a degree exponent
of $3$. When $p=0$, the network is a large-world network and the diameter,
i.e., the largest distance between any two nodes in the system,
increases according to a power law.
When $p=1$, the network is a small-world network and the diameter
is proportional to $\ln N$, where $N$ is the system size.
The phase transition pattern of the Ising model based on this hierarchical
model indicates that as $p$ increases and crosses $p=p^*\cong 0.494$,
the network undergoes a transition from being a large-world network
to being a small-world network \cite{Hinczewski06}.
The transition behavior is different across $p^*$. In the large-world and small-world,
the transition is of the Ising type and the inverted Berezinskii-Kosteritz-Thouless,
respectively.

The density of the permanent cooperators $\rho_c$ is shown as a
function of the temptation payoff $b$ for the two extreme cases corresponding
to $p=0$ and $1$ in \FIG\ref{fig:coop}. While the mean field theory
predicts that no cooperator is present in the PD game for $b>1$, when $p=1$, however,
$\rho_c$ is almost equal to 1 for $b$ values up to $b_c$, beyond which
$\rho_c$ drops suddenly. When $p=0$, $\rho_c$ is relatively small even when
$b=1$, and it decays continuously with increasing $b$. The discontinuity at $b=2$
is because of the deterministic structure of the hierarchical network;
however, such behavior does not appear in other real-world networks.
In fact, such different types of behaviors of $\rho_c$ with respect to $b$ were
also observed in real-world networks, in particular, an email network and a social
network by using the ``Pretty-Good-Privacy" program~\cite{Lozano-Luthi}. In the
former (latter) case, the cooperator density decreases smoothly (drastically).
Furthermore, it is noteworthy that for all $b$, $\rho_c$ at $p=1$ is larger
than $\rho_c$ at $p=0$. This indicates that the shortcuts connecting the
hubs play an important role in enhancing cooperation~\cite{Santos-Duran,Gomez-Poncela}.
The cooperation between influential individuals enhances the overall cooperation
in the society. Let us now proceed to evaluate the case in which $p$ increases from 0 to 1.

\begin{figure}
\begin{center}
\includegraphics[height=8cm,width=8cm]{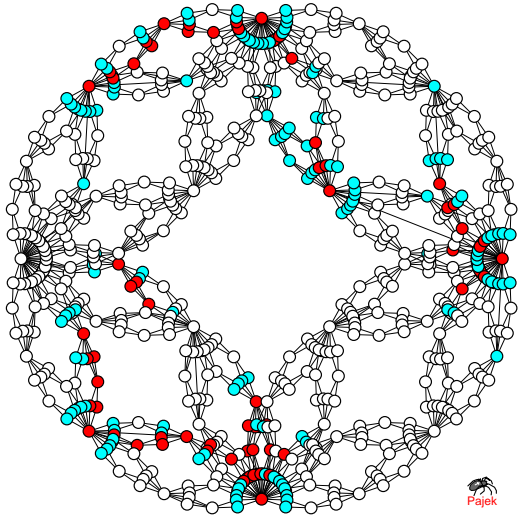}
\caption{(Color online) A snapshot of permanent cooperators (red, dark), permanent
defectors (white), and unstable players (cyan, gray) in the hierarchical
network with $p=0.15$, $b=2.3$ and $N=684$ after 20,000 rounds in a steady
state.} \label{fig:snap}
\end{center}
\end{figure}

A snapshot of the states of the players in the hierarchical network
with $p=0.15$ $b=2.3$ and $N=684$ is shown in \FIG\ref{fig:snap}:
this network represents the large-world network.
The permanent cooperators tend to be located around or at hubs, and may impact
other cooperators. This is because once cooperators
form a cluster around a hub, then the hub becomes stable and is protectd
against any invasion by defectors. However, when defectors gather around the hub,
the hub may cease to be stable because mutual defectors get no payoff.
Therefore, permanent defectors are located at nodes with small degree, while
the unstable players are in between nodes with small degrees and those with large
degrees. We remark that cooperators are not {\it always} located at or
around hubs as they were in the snapshot, because the formation of cooperator
clusters is determined stochastically and it generally depends on the fluctuations
in cooperator densities. The degree of the hubs and the dynamic stochastic
process of the PD game result in the formation of cooperator clusters with a wide
range of sizes.

In contrast, in the small-world network with a large $p$, for example,
$p=1$, permanent cooperators are always located at hubs and mutual cooperation
occurs on a global scale. This behavior occurs even when a hub and
its neighbors are mostly defectors at an early stage; the hub
eventually cooperates with another hub.

The size $G(p,b)$ of the largest cluster of permanent
cooperators is shown as a function of $p$ and $b$ in \FIG\ref{fig:op}(a).
Here $G(p,b)$ is considered to be the order parameter, as in percolation theory.
As $p$ increases for a fixed $b < b_t$ ($b_t$ is defined below),
$G$ increases gradually from 0 to 1 and exhibits a percolation transition at $p_c(b)$.
In Fig.~\ref{fig:op}(a), we can see that the transition interval
in which $0.1 < G < 0.4$ is wide for small $b$.
However, this interval becomes narrower as $b$ increases, indicating that
the giant cluster grows suddenly as $p$ increases for large values of $b$.
A similar behavior is exhibited by the susceptibility,
defined by $\chi(p,b)=\sum_s^{\prime} s^2 n_s$, where the prime
denotes the exclusion of the giant cluster in the summation, and
$n_s$ is the normalized number of $s$-sized clusters composed of
permanent cooperators. As shown in \FIG\ref{fig:op}(b),
the peak positions of $\chi$ mark the phase boundary $p_c(b)$ across which
the giant cooperator cluster increases to a finite size.
The peak height of the susceptibility decreases as $b$ approaches
the tricritical-like point $b_t$, beyond which the susceptibility is likely
to disappear. It is noteworthy that while the peak diverges
in the classical percolation transition, it is finite in the
infinite-order percolation transition, which often occurs in growing
networks~\cite{callaway}. On the basis of this observation, we may say that
the percolation transition is continuous for $p < p_t$ and of
infinite-order for $p > p_t$. The estimated value of $p_t$ is about $0.4$;
this is roughly equal to $p^*\simeq 0.494$, which is
the boundary between the large-world and the small-world networks
as determined on the basis of the thermal transition patterns of the Ising model
\cite{Hinczewski06}.

\begin{figure}
\begin{center}
\includegraphics[width=9cm]{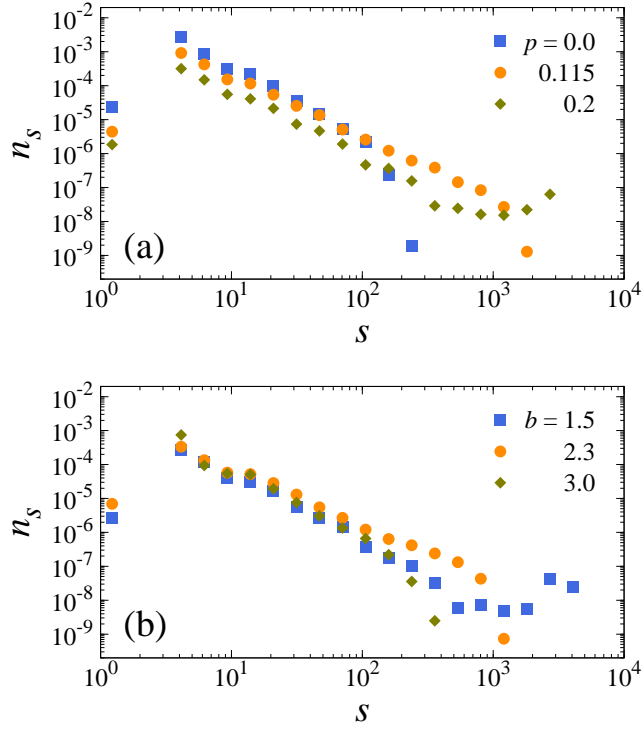}
\caption{(Color online) Size distribution of cooperator clusters in the hierarchical
network with $N=10,924$ nodes for (a) various values of $p$ and $b=1.7$ and (b)
various values of $b$ and $p=0.15$. The data are obtained by averaging the results
for over 500 configurations.}
\label{fig:n_s}
\end{center}
\end{figure}

\begin{figure}
\begin{center}
\includegraphics[width=9cm]{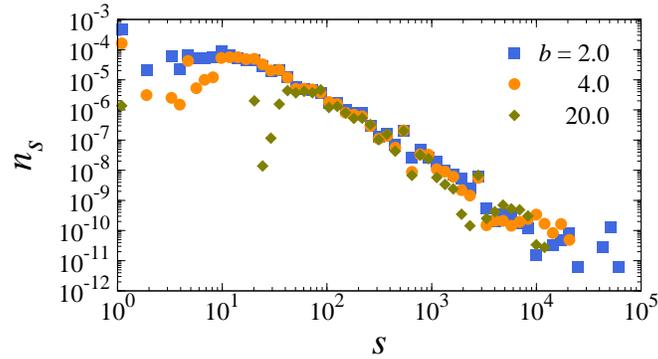}
\caption{(Color online) Size distribution of cooperator
clusters for several $b$ in the the WWW with system size $N=325,729$.
The data are obtained by averaging the results for over 100 configurations.}
\label{fig:n_s_real}
\end{center}
\end{figure}

The cluster-size distribution $n_s$ is shown in
Fig.~\ref{fig:n_s}. In Fig.~\ref{fig:n_s}(a), $n_s$ is plotted against $s$ for several $p$,
but with a fixed $b=1.7$ , and in Fig.~\ref{fig:n_s}(b), $n_s$ is plotted for several $b$,
but with a fixed $p=0.15$. For the large-world network, i.e., the network for which
$p=0.0 < p_c$, $n_s(p)$ exhibits a sub-critical behavior, i.e., for small $s$, it
decays according to a power-law and for large $s$, it decays exponentially beyond a
cutoff. At $p_c\cong 0.1$, $n_s(p_c)$ obeys the power law $n_s(p_c)\sim s^{-\tau}$ with
$\tau \approx 1.85\pm 0.1$. When $p > p_c$, $n_s(p)$ exhibits super-critical behavior.
For the small-world network, the cluster size distribution does not follow a power law.
Similar behavior is observed when $p$ is kept constant while varying $b$ (Fig.~\ref{fig:n_s}(b)). When $p < p_t$, the order parameter increases gradually with $b$, while when $p > p_t$, it increases very drastically.

Large-world networks are known as fractal networks. Typical fractal networks in the real world are the protein interaction network and the World-Wide Web (WWW) \cite{fractal nets}.
In this study, we simulate the evolutionary PD game on the WWW.
Since the WWW network is a single network, corresponding to $p$ being fixed,
we vary only $b$.
As shown in Fig.~\ref{fig:n_s_real}, the distribution of cooperator
cluster sizes for several values of $b$ decays according to a power law
with an exponent of approximately 2. Thus, we conclude that the critical
behavior of the cluster size distribution is not limited
to the hierarchical networks, but rather, this behavior is intrinsic.

In summary, we have studied the percolation transition of cooperator
clusters in artificial hierarchical networks as well as in the WWW network.
We found that in the WWW, the cluster size distribution of permanent cooperators
follows a power law near the percolation threshold.
Such a critical behavior is also observed in the fractal hierarchical networks;
on the basis of these observation, we can determine the condition under which
critical behavior is observed. The power-law behavior indicates
that cooperators create communities of diverse sizes and
at scattered locations. These clusters stochastically form affected by hubs.
In order to improve cooperations on a global scale in society,
communication channels must be established between influential individuals.

This study was supported by an NRF grant (Grant No. 2010-0015066 (BK)),
by the NAP of KRCF (CKY), and by Grants-in-Aid for Scientific Research
(Nos. 20760258 and 20540382, and Innovative Areas
``Systems Molecular Ethology'') from MEXT, Japan (NM).

\end{document}